\documentstyle[prl,aps,preprint,epsfig]{revtex}

\draft

\author{S. Hegyi}

\title{Renormalization group approach to multiparticle density fluctuations}

\address{Research Institute for Particle and Nuclear Physics of the
	Hungarian Academy of Sciences, \\ H-1525 Budapest 114, P.O. Box 49.
	Hungary}

\date{\today}

\begin{document}

\pagestyle{plain}
\pagenumbering{arabic}

\maketitle

\begin{abstract}
  An iterative procedure is developed with the aim of constructing
  homogeneity rules for the distribution $P(\rho,\delta)$ of the particle
  density~$\rho$ at resolution scale~$\delta$. A single iteration step
  consists of a change in the normalization point of $P(\rho,\delta)$
  followed by a rescaling. Similar transformation rule is introduced
  for density fluctuations contaminated by Poisson noise. Application
  of the iterative procedure is given for the Ginzburg-Landau description
  of phase-transition from the quark-gluon plasma and for random cascading
  models.
\end{abstract}


Over the past ten years significant changes have occured in our
knowledge of how to analyse the multiparticle density fluctuations
and correlations, see Ref.~\cite{DDK} for a recent review.
Several new techniques have been developed
dominantly with the same goal: to search for self-similar patterns
possessing no characteristic scale over a certain range of
resolving power. Such fluctuations are expected to
occur e.g. in elementary collision processes due to scale-invariant
branching of parton cascades or in heavy-ion collisions as remnants
of the phase transition from quark-gluon plasma to hadronic matter.

Self-similarity manifests itself in the power-law dependence of
quantities characterizing the strength of fluctuations or correlations
as the resolution scale (bin size in phase-space) is
varied~\cite{DDK}.
Besides power laws,  there is another possible source of scaling
behavior in multiparticle production. The probability distribution
$P(\rho,\delta)$ of the particle density~$\rho$ at resolution
scale~$\delta$ can be a generalized homogeneous function of its
arguments, $P(\rho,\delta)\sim\delta^{-\beta}{\cal F}(u=\rho/\delta^\nu)$
with $\beta$ and $\nu$ being certain parameters~[1]. In this case
scaling refers to the data collapsing feature of homogeneous functions:
after suitable rescaling the data points corresponding to different
distributions $P(\rho,\delta)$ collapse onto a single curve,
${\cal F}(u)$.

Homogeneity rules play a central role in the theory of critical
phenomena.
Near the critical point of a physical system the
thermodynamic functions exhibit homogeneous form~\cite{Widom}
which implies the
existence of a law of corresponding states. Using a suitably
chosen scaling transformation it is possible to bring different
states of the same system to coincidence and thus to compress
many experimental or theoretical results into a compact form.
This scaling hypothesis was first justified by the
spin decimation procedure~\cite{Kadan},
the prototype of renormalization
group (RG) transformations in statistical
physics~\cite{WilKo,Ma}.

The connection between scale-invariant density fluctuations
in multiparticle dynamics and the scaling laws of statistical physics
has already been established~\cite{QGP1,Viet}.
In this Letter an attempt is made
to deepen the relationship between the two fields
by developing an iterative procedure that acts
on the distribution $P(\rho,\delta)$ and provides the analogue
of RG transformations. In the course of the iteration one may (and will)
arrive at scaling behavior of $P(\rho,\delta)$ even if the
distribution is not a homogeneous function of its arguments
at the beginning of the procedure.

Let us start with some elementary properties of the
probability distribution  $P(\rho,\delta)$
of observing particle density~$\rho$ at
resolution scale (bin size)~$\delta$. For simplicity we shall
consider here only averaging over the events. The effect of statistical
noise will be discussed later on. The probability density
$P(\rho,\delta)$ fulfills the normalization condition
	$\int_0^\infty P(\rho,\delta)d\rho=1$.
Its moments are defined by
	$\langle\rho^q(\delta)\rangle=
	\int_0^\infty\rho^q P(\rho,\delta)d\rho$.
If only the first moment
	$\langle\rho(\delta)\rangle$
is responsible for the $\delta$-dependence of $P(\rho,\delta)$
the distribution can be written in the homogeneous form
\begin{equation}
	P(\rho,\delta)=\frac{1}{\langle\rho(\delta)\rangle}\,
	{\cal F}\!\left(\frac{\rho}{\langle\rho(\delta)\rangle}\right)
	\label{Scal_0}
\end{equation}
where the function~${\cal F}$ depends only on the scaling variable
	$u=\rho/\langle\rho(\delta)\rangle$.
It satisfies the two normalization conditions
	$\int_0^\infty{\cal F}(u)du=
 	\int_0^\infty u{\cal F}(u)du=1$.
Thus we have two properly normalized scaling functions, the second one is
	$\widetilde{\cal F}(u)=u{\cal F}(u)=\rho P(u)$.
Obviously, the normalized moments of $P(\rho,\delta)$ coincide with the
moments of ${\cal F}(u)$,
\begin{equation}
	\frac{\langle\rho^q(\delta)\rangle}
	{\langle\rho(\delta)\rangle^q}=
	\int_0^\infty u^q{\cal F}(u)du=C_q
\end{equation}
which are independent of~$\delta$. Moreover, the moments of
$\widetilde{\cal F}(u)$ are given by
	$\widetilde C_q=\int_0^\infty u^q\,\widetilde{\cal F}(u)du
	=C_{q+1}$
i.e. the difference between the two scaling functions amounts to a
shift in the ranks of their moments.

Although the scaling law of type Eq.\ (\ref{Scal_0}) is frequently
encountered in nature, for self-similar multiparticle density
fluctuations it can not be valid. The observations indicate power-law
dependence of the normalized moments of $P(\rho,\delta)$ as the
resolution scale is varied,
	$C_q(\delta)\propto\delta^{-\varphi_q}$
where the $\varphi_q$ are the intermittency exponents~\cite{DDK}.
Accordingly, we have \hbox{$\delta$-dependent} ``scaling'' functions
${\cal F}(u,\delta)$ and $\widetilde{\cal F}(u,\delta)$.
Let us observe that the latter enables one
to perform a rescaling of type Eq.\ (\ref{Scal_0}) yet again,
the new scaling hypothesis reads
\begin{equation}
	\widetilde{\cal F}(u,\delta)=\frac{1}{\widetilde C_1(\delta)}\,
	{\cal F}_1\!\left(\frac{u}{\widetilde C_1(\delta)}\right).
	\label{Scal_1}
\end{equation}
(Rescaling ${\cal F}(u,\delta)$ in a similar manner has no effect
since $C_1(\delta)=1$.) The subscript of the new scaling function
refers to the first step of an iterative sequence of transformations
acting on the distribution $P(\rho,\delta)$. For the starting
distributions and moments corresponding to the $0$th step
we shall not indicate this explicitly.

Our new scaling variable is
	$u_1=u/\widetilde C_1(\delta)$
and the first-iterate moments
	$C_{q,1}=\int_0^\infty u_1^q\,{\cal F}_1(u_1)du_1$
are related to the original ones through
\begin{equation}
	C_{q,1}=\frac{\widetilde C_q(\delta)}
	{[\widetilde C_1(\delta)]^q}=
	\frac{C_{q+1}(\delta)}{[C_2(\delta)]^q}.
	\label{C_q,1}
\end{equation}
Expressing now the scaling function ${\cal F}_1(u_1)$ similarly
we get
	${\cal F}_1(u_1)=C_2(\delta)\,\rho\,
	P\,\big[u/C_2(\delta)\big]$
which can be rewritten as
\begin{equation}
	{\cal F}_1(u_1)=
	\frac{\langle\rho^2(\delta)\rangle}
	{\langle\rho(\delta)\rangle}
	\frac{\rho}{\langle\rho(\delta)\rangle}\,
	P\left(\rho\left/
	\frac{\langle\rho^2(\delta)\rangle}
	{\langle\rho(\delta)\rangle}\right.\right).
\end{equation}
It is seen that ${\cal F}_1(u_1)$ has the same form as ${\cal F}(u)$,
namely,
	${\cal F}_1(u_1)=\langle\rho(\delta)\rangle_1\,
	P_1\big[\rho/\langle\rho(\delta)\rangle_1\big]$
where the first-iterate distribution $P_1(\rho,\delta)$
and its moments $\langle\rho^q(\delta)\rangle_1$
are obtained by changing the normalization point
of $P(\rho,\delta)$ from the $0$th moment
to $\langle\rho(\delta)\rangle$ and
rescaling by $\langle\rho(\delta)\rangle$ to maintain the
overall normalization.

Demonstrating the benefits of the above transformation of
$P(\rho,\delta)$ let us assume that the self-similar density
fluctuations show monofractal properties. Then the intermittency
exponents are given by $\varphi_q=\varphi_2(q-1)$ and the anomalous
fractal dimensions $D_q=\varphi_q/(q-1)$ are $q$-independent, $D_q=D_2$.
The normalized moments of $P(\rho,\delta)$ take the form
	$C_q(\delta)=A_q[C_2(\delta)]^{q-1}$
with coefficients $A_q$ independent of~$\delta$. However,
Eq.\ (\ref{C_q,1}) yields
	$A_q=C_{q-1,1}$.
This implies that monofractal density fluctuations give rise not only
to power-law scaling of the normalized moments of $P(\rho,\delta)$
but also to homogeneous form of the distribution after changing its
normalization point to $\langle\rho(\delta)\rangle$ and
making an appropriate rescaling.

In the followings we omit explicit reference to the
variable~$\delta$ to simplify the notation. If ${\cal F}_1(u_1)$
is not independent of the resolution scale, the single
iteration step described before can be repeated until the appearance
of scaling (i.e. data collapsing) behavior provided that certain
moments of $P(\rho)$ exist, see below. In the $k$th step of the
iteration we have the normalization conditions
	$\int_0^\infty{\cal F}_k(u_k)du_k=
	\int_0^\infty\widetilde{\cal F}_k(u_k)du_k=1$
where
	$\widetilde{\cal F}_k(u_k)=u_k\,{\cal F}_k(u_k)$.
The distribution
	${\cal F}_k(u_k)=
	\langle\rho\rangle_k\,P_k\big(\rho/\langle\rho\rangle_k\big)$
is related to the $(k-1)$th step through
\begin{equation}
	{\cal F}_k(u_k)=\widetilde C_{1,k-1}\;\widetilde{\cal F}_{k-1}\!
	\left(\frac{u_{k-1}}{\widetilde C_{1,k-1}}\right)
\end{equation}
and its moments
	$C_{q,k}=\int_0^\infty u_k^q\,{\cal F}_k(u_k)du_k$
are given by
\begin{equation}
	C_{q,k}=\frac{\widetilde C_{q,k-1}}{\widetilde C_{1,k-1}^q}=
	\frac{C_{q+1,k-1}}{C_{2,k-1}^q}.
\end{equation}
The $(k+1)$th step of the iteration follows in the same way as the
first step in Eq.\ (\ref{Scal_1}),
\begin{equation}
	\widetilde{\cal F}_k(u_k)=\frac{1}{\widetilde C_{1,k}}\;
	{\cal F}_{k+1}\!\left(\frac{u_k}{\widetilde C_{1,k}}\right).
\end{equation}
Scaling behavior of $P_{k+1}(\rho)$ appears if the monofractal
type relationship $C_{q,k}\propto C_{2,k}^{q-1}$ holds
with constants of proportionality independent of the resolution
scale~$\delta$.

Let us now consider how the distribution $P_k(\rho)$ is related to the
original $P(\rho)$.
The connection is provided by the straightforward extension
of the normalization point changing transformation discussed previously,
\begin{equation}
	P_k(\rho)=\frac{\rho^k}{\langle\rho^k\rangle}
	P(\rho)\label{P_n,k}
\end{equation}
and thus the moments of $P_k(\rho)$ are given by
	$\langle\rho^q\rangle_k=
	\langle\rho^{q+k}\rangle/\langle\rho^k\rangle$. From
this we get for the normalized moments $C_{q,k}$ the expression
\begin{equation}
	C_{q,k}=\frac{\langle\rho^q\rangle_k}{\langle\rho\rangle_k^q}=
	C_{q+k}\frac{C_k^{q-1}}{C_{k+1}^q}
\end{equation}
in terms of the original normalized moments $C_q$.
Homogeneity of $P_{k+1}(\rho)$ appears in case of
$\delta$-independence of the product
	$C_{k+q}C_{k+2}^{1-q}C_{k+1}^{q-2}$.
The necessary
condition of performing the $k$th iteration step is the existence of
the moments of $P(\rho)$ up to $k$th order.

Eq.\ (\ref{P_n,k}) is analogous to the
RG transformations which correspond in many important cases to a change
in the norm of the parameters characterizing a physical system~\cite{Ma}.
It is a \hbox{Gell-Mann} and Low type relationship~\cite{GML} in which a
multiplicative transformation (in our case multiplication by $\rho^k$)
is compensated by a rescaling and a suitable change in the
renormalized parameters. Since the operation
	$P_k(\rho)\to P_{k-1}(\rho)$
does not exist the iteration steps constitute a semi-group. In
statistical physics the Kadanoff-Wilson type RG
transformations perform a systematic
reduction in the number of degrees of freedom (spin decimation).
This enables one to eliminate the small-scale fluctuations from
the problem which are irrelevant to critical point behavior
such as the homogeneity of thermodynamic functions.
In the iterative procedure developed here the elimination of
fluctuations manifests through elimination of the moments of $P(\rho)$
via moment-shifting and rescaling until one observes a homogeneity
rule of type Eq.\ (\ref{Scal_0}) for $P_k(\rho)$.

The finite number of detected particles gives rise to statistical
fluctuations contaminating
the distribution $P(\rho)$ of purely dynamical origin.
It is convenient
to represent the statistical component by a Poisson process whose
convolution with $P(\rho)$ can be described by the Poisson
transform~\cite{DDK}.
It yields the discrete probability distribution $P(n)$ of detecting
$n$ particles:
\begin{equation}
	P(n)=\int_0^\infty P(\rho)\frac{\rho^n}{n!}e^{-\rho}d\rho.
\end{equation}
The Poisson transform has the useful property that the factorial
moments $\xi_q$ of $P(n)$ are equivalent to
the ordinary moments of $P(\rho)$,
\begin{equation}
	\xi_q=\langle n^{[q]}\rangle=
	\sum_{n=q}^\infty n^{[q]}P(n)=
	\langle\rho^q\rangle
\end{equation}
where
	$n^{[q]}=n(n-1)\ldots(n-q+1)$.
Thus in the presence of Poisson noise
factorial observables make it possible to study directly the
dynamical fluctuations whose self-similarity is signalled by the
power-law dependence of the normalized factorial moments
$F_q=\xi_q/\xi_1^q$ on the resolution scale~$\delta$~\cite{BP}.

We are interested here in the Poisson transform of the iterated dynamical
distribution $P_k(\rho)$. But let us first consider briefly some
properties of Eq.\ (\ref{P_n,k}) itself. This transformation of
$P(\rho)$ is known as size-biasing of order $k$ in the mathematical
literature~\cite{Ord}.
A worth noting feature is that the log-exponential
family of continuous probability densities are form-invariant under
size-biasing, only their parameters are affected. In this respect the
log-normal distribution deserves special attention. Its parameters are
changed by size-biasing of order $k$ according to
	$\mu\to\mu+k\sigma^2$
with $\mu$ and $\sigma^2$ being the mean and variance of the original
distribution. Therefore a log-normal $P(\rho)$ is a fixed point of our
iterative procedure: although Eq.\ (\ref{P_n,k}) changes its first
moment, this change is scaled out by constructing ${\cal F}_k(u_k)$ and
we get an unaffected scaling function at each iteration step.

The Poisson transform of $P_k(\rho)$ leads to size-biasing of factorial
order $k$ acting on $P(n)$:
\begin{equation}
	P_k(n)=\int_0^\infty P_k(\rho)\frac{\rho^n}{n!}e^{-\rho}d\rho=
	\left.\frac{n^{[k]}}{\langle n^{[k]}\rangle}P(n)
	\right|_{\Leftarrow k}
\end{equation}
where the symbol $\Leftarrow$$k$ denotes left-shifting
the size-biased $P(n)$ by $k$ places.
It is seen that we have the same kind of transformation as for
$P(\rho)$: the normalization
point of $P(n)$ is changed to $\langle n^{[k]}\rangle$ and a rescaling
by $\langle n^{[k]}\rangle$ is performed to maintain the overall
normalization. This yields a discrete
distribution with $k$ leading zeroes,
\hbox{$P(n<k)=0$}, hence a left-shift by
$k$ places is required to maintain also the original support in $n$
of the nonzero probabilities~\cite{Ord}.
The normalized factorial moments
$F_{q,k}$ are related to $F_q$ in the same way as
$C_{q,k}$ to $C_q$ i.e.
	$F_{q,k}=F_{q+k}F_k^{q-1}F_{k+1}^{-q}$.
The expression for $P_k(n)$ implies that the
normalization point changing transformations form
a discrete semi-group in the presence of Poisson noise.
On the contrary, $P_k(\rho)$ is well defined
for fractional $k$ too although the iterative procedure that led us
to Eq.\ (\ref{P_n,k}) enables only integer valued~$k$.
It is worth mentioning that $F_{2,k-2}=F_kF_{k-2}F_{k-1}^{-2}$
yields for $k\geq2$ the $k$th bunching parameter~\cite{Bunch}
in the limit $\delta\to0$.

We now turn our attention to applications of our results.
In recent years considerable interest has been devoted to the
Ginzburg-Landau theory of the phase transition from quark-gluon plasma
to hadronic matter~\cite{Hwa1,Hwa2}.
Instead of a strict power-law scaling
$F_q\propto\delta^{-\varphi_q}$ of the normalized factorial moments,
a so-called $F$-scaling behavior
$F_q\propto F_2^{\beta_q}$ is observed with $\beta_q=(q-1)^\nu$.
For second-order transition $\nu=1.304$ and in the case of a first-order
transition $\nu=1.45$. The $F$-scaling rule of the above form has the
property $F_0\neq1$ hence it is difficult to identify the
family of discrete
probability laws that belongs to the observed $F_q$-pattern.
Performing a single
iteration step the factorial moments change according to
$F_{q,1}\propto F_2^{\beta_q}$ with $\beta_q=q^\nu-q$ and therefore
\begin{equation}
	F_{q,1}\propto F_{2,1}^{\beta_q}\quad\mbox{with}\quad
	\beta_q=\frac{q^\nu-q}{2^\nu-2}\label{Levy}.
\end{equation}
This form of $F$-scaling is the log-L\'evy law~\cite{Levy} with
L\'evy index $0<\nu\leq2$. We have thus identified the family of
probability laws for $P_1(n)$ in the Ginzburg-Landau formalism
with a strict bound on~$\nu$.
The theoretically relevant values are within the allowed range.
Our result can be checked in a laser experiment at the threshold of
lasing where the photocount fluctuations are in accordance with the
 Ginzburg-Landau theory of second-order phase transitions and the
$F$-scaling rule with $\nu=1.304$ has been experimentally
confirmed~\cite{Hwa3}.

The log-L\'evy law $F_q\propto F_2^{\beta_q}$ with $\beta_q$ given
in Eq.\ (\ref{Levy}) is introduced to describe multiplicative
random cascading models~\cite{Levy}.
Let us keep track of the evolution of factorial moments
in the course of successive iteration steps. It can be characterized
by the ratio
\begin{equation}
	\frac{D_{q,k}}{D_q}=
	\frac{(q+k)^\nu+(q-1)k^\nu-q(k+1)^\nu}{q^\nu-q}
\end{equation}
whose $k$-dependence is shown in Fig.~1 for $q=2$ and for
different values of the L\'evy index~$\nu$. Smaller is the value of $\nu$\
faster is the convergence to $D_{2,k}=0$, i.e. to
$\delta$-independence of $F_{2,k}$. This is what we expect since $\nu=0$
corresponds to monofractal behavior which yields $\delta$-independent
factorial moments after a single iteration step.
The $\nu=2$ limiting case
of the L\'evy index defines maximum degree of multifractality.
Random cascades of this type are described by
the log-normal law whose fixed point property
discussed before is apparent from Fig.~1.
For $\nu>2$, which is not permitted for log-L\'evy laws,
$F_{2,k}$ gets farther away from $\delta$-independence.

Our last example concerns the analogy between fractal structures
and statistical mechanics~\cite{DDK,Pes}.
In the thermodynamic description of multifractal cascades
the rank~$q$ plays the role of inverse temperature~\cite{Pes}.
Analogously to RG calculations in statistical physics, performing
the iterative sequence of transformations can be thought as
varying the temperature. The evolution of
$D_{q,k}$ allows the introduction of an effective inverse temperature
$q_k$ whose variation with $k$ can be obtained from
the equation
	$(q+k)^\nu+(q-1)k^\nu-q(k+1)^\nu=q_k^\nu-q_k^{\ }$.
The L\'evy index~$\nu$ controls the rate at which the effective
inverse temperature changes in the course of the iteration. Therefore
$\nu$ plays the role of the critical exponent of the
correlation length characterizing condensed matter systems
at the critical temperature~\cite{Ma}.

In conclusion, we have developed an iterative procedure capable of
finding homogeneity rules for a distribution that describes the
fluctuation properties of a certain physical system. A single iteration
step consists of a change in the normalization point
of the distribution compensated by
a suitable change in the renormalized parameters and a rescaling.
The number of steps
required to arrive at data collapsing behavior measures how strongly
the scaling hypothesis of type Eq.\ (\ref{Scal_0}) is violated.
For self-similar fluctuations it is maximal in the case of
maximum degree of
multifractality, providing the fixed point of the iteration, and minimal
for monofractals which give rise to scaling behavior after a single
iteration step. Concerning multiparticle dynamics,
our results shed new light on how to characterize
the degree of KNO scaling violation
of the full phase-space multiplicity distributions~\cite{KNO}
in terms of novel scaling laws.
Since homogeneity rules and data collapsing deserve special
attention in many areas of physics the iterative procedure
can have a number of applications in other
fields as well.

\bigskip

This research was supported by the Hungarian Science Foundation
under grant OTKA-F4019/1992.

\newpage

\begin{figure}
\vskip2cm
\hskip1.7cm\epsfig{file=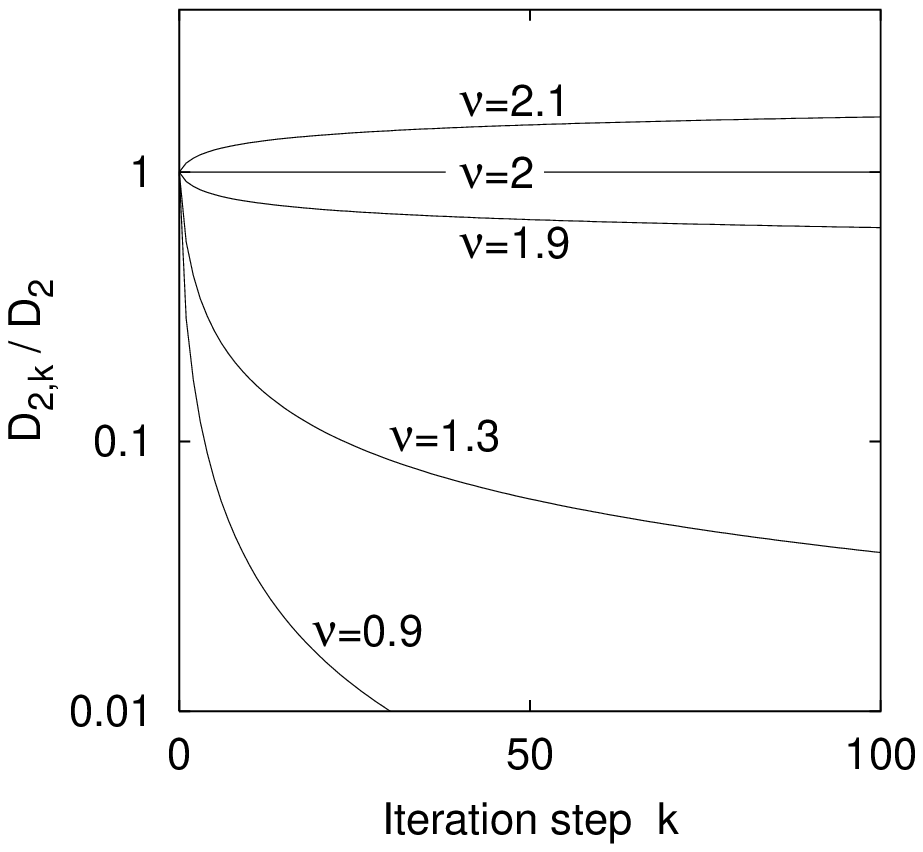}
\bigskip
\caption{Variation of the anomalous fractal dimension $D_2$
with iteration step $k$
for different values of the L\'evy index $\nu$.}
\end{figure}

\end{document}